\documentclass[12pt,preprint]{aastex}

\slugcomment{Accepted for publication in Astrophys. J. Lett.}

\shorttitle{Probing Heliospheric Asymmetries}
\shortauthors{Pogorelov, Heerikhuisen \& Zank}

\begin{document}


\title{Probing Heliospheric Asymmetries with an MHD-kinetic model}


\author{Nikolai V. Pogorelov, Jacob Heerikhuisen, Gary~P. Zank}
\affil{Institute of Geophysics and Planetary Physics, University of California,
    Riverside, CA 92521, U.S.A.}




\begin{abstract}
New solar wind data from the \textit{Voyager}~1 and \textit{Voyager}~2 spacecraft, together with the \textit{SOHO} SWAN measurements of the
direction that neutral hydrogen enters into the inner heliosheath
and neutral helium measurements provided by multiple observations are expected to provide
more reliable constraints on the ionization ratio of the local interstellar medium (LISM)
and the direction and magnitude of the interstellar magnetic field (ISMF).
In this paper we use currently the most sophisticated numerical model of the heliospheric
interface, which is based on an MHD treatment of the ion flow and kinetic modeling of neutral particles, to
analyze an ISMF-induced asymmetry of the heliosphere in the presence of the interplanetary magnetic field
and neutral particles. It is shown that secondary hydrogen atoms modify the LISM properties leading to its shock-free
deceleration at the heliopause.
We determine the deflection of hydrogen atoms
from their original trajectory in the unperturbed LISM and show that it occurs not only in the plane defined by the
ISMF and LISM velocity vectors, but also, to a lesser extent, perpendicular to this plane.
We also consider the possibility of using 2--3 kHz radio emission data to further constrain the ISMF direction.

\end{abstract}


\keywords{ISM: kinematics and dynamics, magnetic fields; Sun: solar wind; physical data and processes: MHD, shocks}

\section{Introduction}
The interaction of the solar wind (SW) with the local interstellar medium (LISM) determines the structure
of the outer heliosphere. The mathematical problem of this interaction should articulate
a physical model that takes into account the fundamental processes occurring in the SW plasma
when it collides with a weakly ionized LISM. The model should be accompanied by boundary conditions
at large distances from the Sun, in the unperturbed LISM, and at 1~AU, where the SW properties are monitored
by a fleet of spacecraft, such as \textit{ACE}, \textit{WIND}, \textit{Ulysses}, \textit{STEREO}, etc.
Ultimately, the solution to the problem of
providing the boundary conditions on an inner surface must be based on the use of solar observations
(see, e.~g., Wang \& Sheeley~2006). The properties of the partially ionized LISM generally cannot be measured \emph{in situ}.
For this reason, we are obliged to rely on numerical modeling, which is capable of providing parameter-dependent results
for the quantities measured in the inner heliosheath.

The direction and magnitude of the LISM velocity $\mathbf{V}_\infty$ are determined
from the observations of He atoms (see M\"obius et al.~2004 and references therein), which can propagate easily
into the SW region within 1~AU from the Sun. The temperature of H ions and atoms in the LISM is usually
assumed to be the same as that of neutral He. The number densities of the LISM H atoms ($n_{\mathrm{H}\infty}$) and ions ($n_\infty$),
and even the ionization ratio of the hydrogen plasma, are not well know. Constraints on the densities can be derived from
the ratio of neutral H and He line densities, which is 14.7 \citep{Dupuis,Wolff}, and the abundance ratio between H
and He as chemical elements in the LISM, which is 10 \citep{Vallerga96}. Several MHD calculations that include neutral particles
self-consistently \citep{Izmod05,Jacob06,Pogo07}
use $n_\mathrm{H\infty}$ in the range 0.15 to 0.2~cm$^{-3}$, which yields a neutral H density of about 0.1~cm$^{-3}$
at the SW termination shock (TS) -- a value consistent with the Ulysses pick-up ion (PUI) measurements \citep{Gloeckler}.

The strength and direction of the interstellar magnetic field (ISMF) are not determined with sufficient accuracy.
According to starlight polarization measurements in the solar neighborhood \citep{Tinbergen},
the ISMF vector $\mathbf{B}_\infty$ belongs to the Galactic plane.
The theory for the origin of the Local Bubble suggested by \citet{Cox03} requires strong
magnetic fields of $4$--$5\,\mu \mathrm{G}$ nearly parallel to the LISM velocity. Recent measurements performed by the
Solar Wind ANisotropy (SWAN) experiment on the \textit{SOHO} spacecraft determined the direction that neutral hydrogen
enters the inner heliosheath \citep{Lallement05}. This direction and the initial direction of the LISM
velocity (the angle between these is $4^\circ \pm 1^\circ$) define a so-called hydrogen deflection plane (HDP).
The deflection of neutral H flow occurs through charge exchange
in the regions where there is a difference between the velocities of H atoms and ions.
Since it is assumed that the velocities of H atoms and He atoms coincide with that of H ions and because the proton density
in the SW is substantially smaller than that at the LISM side of the heliopause (HP), the major deflection takes place
in the outer heliosheath -- the region of compressed plasma on the interstellar side of the~HP. Thus, it is reasonable to assume that
the deflection of the neutral H flow is caused by an asymmetry in the LISM plasma distribution just outside the HP.
This suggestion of \citet{Lallement05} was confirmed
by \citet{Izmod05} who simulated the SW--LISM interaction with an ISMF vector $\mathbf{B}_\infty$ lying in
the HDP at $45^\circ$ to $\mathbf{V}_\infty$. The asymmetry in the plasma distribution caused by
the obliquity of the ISMF (see Pogorelov \& Matsuda~1998; Ratkiewicz et al.~1998; Pogorelov et al.~2004, 2006) does indeed
result in an H-atom deflection angle close to that determined by \citet{Lallement05}.
If we ignore the interplanetary magnetic field (IMF), as it was done by \citet{Izmod05},
the plane defined by the vectors $\mathbf{B}_\infty$ and $\mathbf{V}_\infty$ ($B$-$V$ plane) is the symmetry plane for the heliosphere.
As a result, the deflection of the neutral H flow in the direction perpendicular to the $B$-$V$ plane should on the average be zero.
However, as shown by \citet{Pozank06} and \citet{Pogo07}, this is not precisely so, since the presence of IMF
eliminates the plane of symmetry and therefore allows an H-atom deflection perpendicular for the $B$-$V$ plane.

With V2 crossing the TS at a distance of about 84~AU (according to the announcement of the \textit{Voyager} science team
at the AGU Fall Meeting 2007), which is 10~AU
closer to the Sun than V1 did in December~2004, it is interesting to analyze the asymmetry of the heliosphere
using a fully three-dimensional numerical model based on ideal MHD description of the interacting SW and LISM ions
that incorporates both the IMF and ISMF \citep{Pozaog04,Pozaog06,Pogo07}, together with a self-consistent kinetic description for H-atoms
\citep{Jacob06,Jacob07}. These two very different sets of equations
are coupled via source terms that describe momentum and energy transfer between atoms and ions through resonant charge
exchange. Only on the basis of a self-consistently coupled model can we determine conclusively whether the ISMF can make the HP sufficiently
asymmetric, so that the TS is substantially closer to the Sun in the V2 direction than in the V1 direction.
This letter addresses the above issues
by analyzing the distribution of charged and neutral particles in the heliosphere for the case when the $B$-$V$ plane is parallel
to the HDP determined by \citet{Lallement05}.
We also discuss the possibility of using 2--3 kHz radio emission data \citep{Kurth03}
to further constrain the ISMF orientation.

\section{ISMF-induced asymmetry of the heliosphere}
We perform calculations in a Cartesian coordinate system with the origin at the Sun. The $x$-axis is oriented along the Sun's rotation axis,
which we assume to be perpendicular to the ecliptic plane. The $z$-axis belongs to the plane defined by the $x$-axis and $\mathbf{V}_\infty$,
and is directed toward the LISM. The $y$-axis completes the right coordinate system. The direction of the LISM velocity
is known to be aligned with the vector $\mathbf{r}_\mathrm{He}=(-0.087156, 0, -0.9962)$. The HDP is defined by $\mathbf{r}_\mathrm{He}$ and
the vector $\mathbf{r}_\mathrm{H}=(-0.1511, -0.04049, -0.9877)$ at which LISM neutrals enter the inner heliosheath, according to
the observations of \citet{Lallement05}. Throughout this paper we shall assume that $\mathbf{B}_\infty$ is aligned with the vector
$\mathbf{r}_B=(-0.5, -0.2678, -0.82356)$. Thus, $\mathbf{B}_\infty$ belongs to the observed HDP
and is directed into the southern hemisphere at an angle of $30^\circ$
to the ecliptic plane. This direction gave one of the largest V1--V2 asymmetries of the TS in the two-fluid
(one ion fluid and one neutral fluid) calculations of \citet{Pogo07}. As noticed by \citet{Opher06} and \citet{Pogo07},
this orientation is consistent with the direction of energetic proton fluxes
observed by V1 and V2. However, the presence of neutrals does not allow V2 be directly connected to the TS at distances greater than 3~AU
\citep{Pogo07}. The LISM plasma velocity, temperature and
density are $V_\infty=26.4\ \mathrm{km}\,\mathrm{s}^{-1}$, $T_\infty=6527\,\mathrm{K}$,
and $n_\infty=0.06\ \mathrm{cm}^{-3}$, respectively.  It is assumed that the SW is spherically symmetric with
the following parameters at 1~AU:
$V_\mathrm{E}=450\,{\rm km\, s^{-1}}$, $T_\mathrm{E}= 51100\,\mathrm{K}$, and $n_\mathrm{E}=7.4\, {\rm cm}^{-3}$.
The density of neutral hydrogen is $n_\mathrm{H\infty}=0.15\ \mathrm{cm}^{-3}$.
The magnitude of the ISMF vector is $B_\infty=3\, \mu\mathrm{G}$. The radial component of the IMF at 1~AU is equal to $37.5\,\mu\mathrm{G}$.
Figure~\ref{front_view} shows the front view of the HP, obtained with our MHD-kinetic model,
cut by the HDP as determined by \citet{Lallement05},
as well as the V1 and V2 trajectories.  We also show the orientation of the Galactic plane.

\begin{figure}
\epsscale{0.9}
\centerline{
\plotone{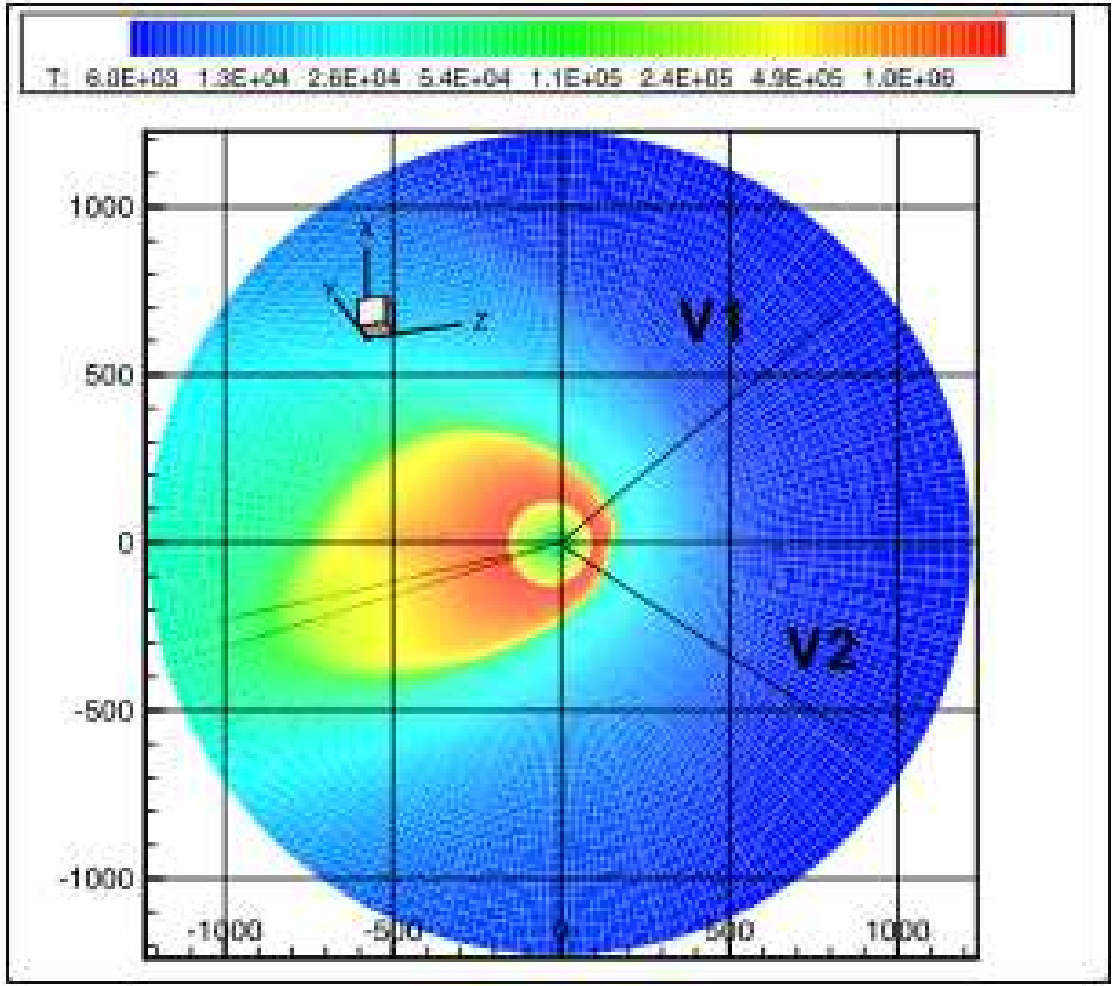}
}
\caption{Plasma temperature distributions the V1--V2 plane. The straight lines
show the V1 and V2 trajectories.}
\label{T-distribution}
\end{figure}

\begin{figure*}
\centerline{
\includegraphics[width=0.25\textwidth]{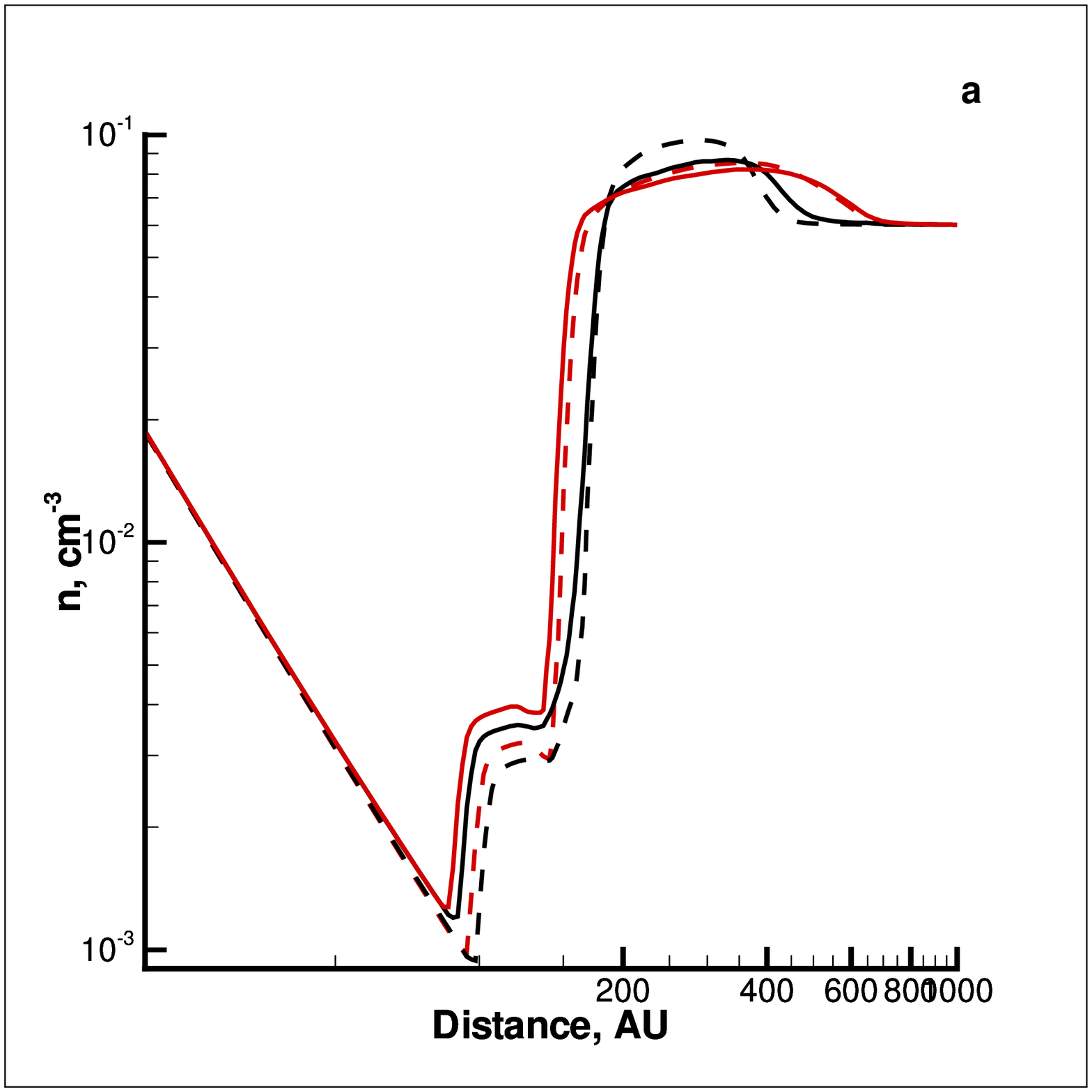}\hspace{20.mm}
\includegraphics[width=0.25\textwidth]{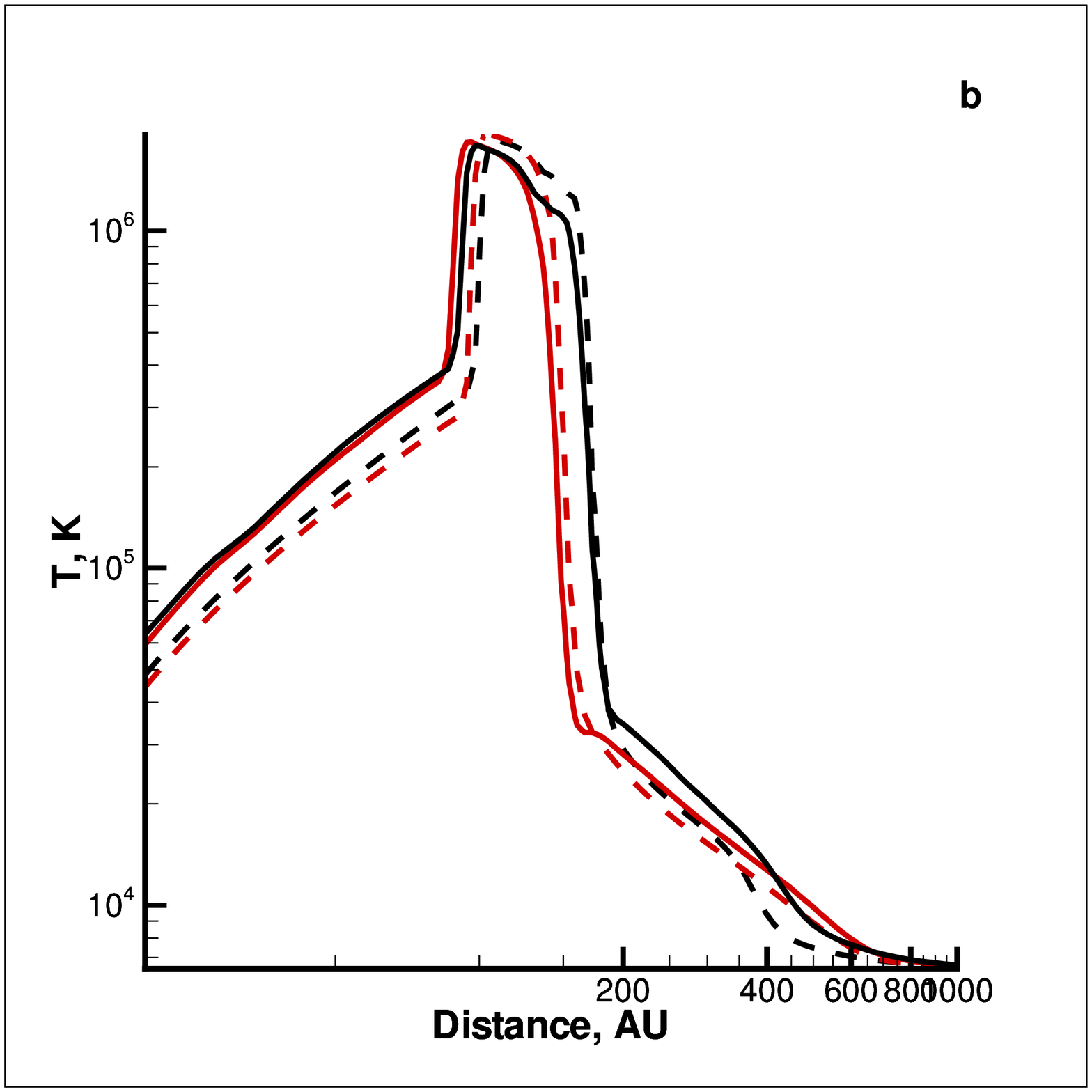}
}
\caption{The distributions of plasma (a) density and (b) temperature in the V1 (black lines) and V2 (red lines) directions. The results
shown with solid and dashed lines are obtained with MHD-kinetic and 5-fluid codes, respectively.}
\label{rho_T}
\end{figure*}

Figure~\ref{T-distribution} shows the distribution of the plasma temperature in the V1--V2 plane.
It is seen that, in agreement with the two-fluid calculations of \citet{Pogo07}, the asymmetry of the TS is minor. This contrasts with
the ideal MHD calculations of \citet{Opher06}, where the absence of neutral atoms tends to exaggerate the asymmetry.
To be more quantitative, in Fig.~\ref{rho_T} we show the distributions of the proton density and
temperature in the directions of V1 (solid black lines) and V2 (solid red lines). Here, for the sake of comparison,
we also use dashed lines to show the same distributions obtained with a 5-fluid model. The latter is based on the solution of ideal MHD
equations to model the flow of protons and four
coupled sets Euler equations to simulate the flow of separate neutral H fluids. These consist of the parent
LISM neutrals (population~0) and those born in the outer heliosheath (population 1), inner heliosheath (population 2), and supersonic
SW (population~3). It is interesting to notice in Fig.~\ref{rho_T} that secondary neutral atoms belonging to populations~1 and~2
modify the parameters of the LISM plasma to such an extent that a bow shock, intrinsic to ideal MHD models, disappears.
The introduction of population~1 neutrals considerably improves the quality of the results in the outer heliosheath
\citep{Jacob06}. However, the discrepancy in the SW temperature is still large because the filtration of neutral H at the HP
is greater in the multi-fluid model than in the MHD-kinetic model. We note here that charge exchange
plays the major role in determining both the geometrical features of the heliospheric interface and the distribution of plasma quantities
inside the heliosphere \citep{Bama93,Zank96,Pozaog06}. Charge exchange decelerates the supersonic SW  and introduces suprathermal PUIs
with a thermal velocity corresponding to the supersonic SW speed. Since we do not distinguish here between the core proton population and PUIs, this results
in the SW proton temperature increase. As shown by \citet{Pogo07}, plasma-neutral collisions reduce
the asymmetry of the heliosphere caused by the pressure exerted by the ISMF in a direction perpendicular to its orientation.
This occurs because an asymmetric distribution of hydrogen ions ahead of the HP leads to enhanced charge exchange in the regions with higher ion
density. This introduces secondary ions that have velocity of the parent LISM neutrals and exert additional pressure counterbalancing
the ISMF pressure effect. It is seen that
the TS is closer to the Sun in the V2 direction than in the V1 direction by about 3~AU only in our MHD-kinetic simulation.
In any event, the steady-state asymmetry is too small to ensure the V2 crossing the TS at a distance to the Sun closer
by 10~AU than V1 \citep{Stone}. Instead, it is likely that temporal variations in the SW ram pressure will modify the TS location
significantly \citep{Barnes,Suess,Klaus,Zank03,Horst,Pogo07,Haru07}, thus supplementing the ISMF-pressure effect.

\begin{figure*}
\centerline{
\includegraphics[width=0.25\textwidth]{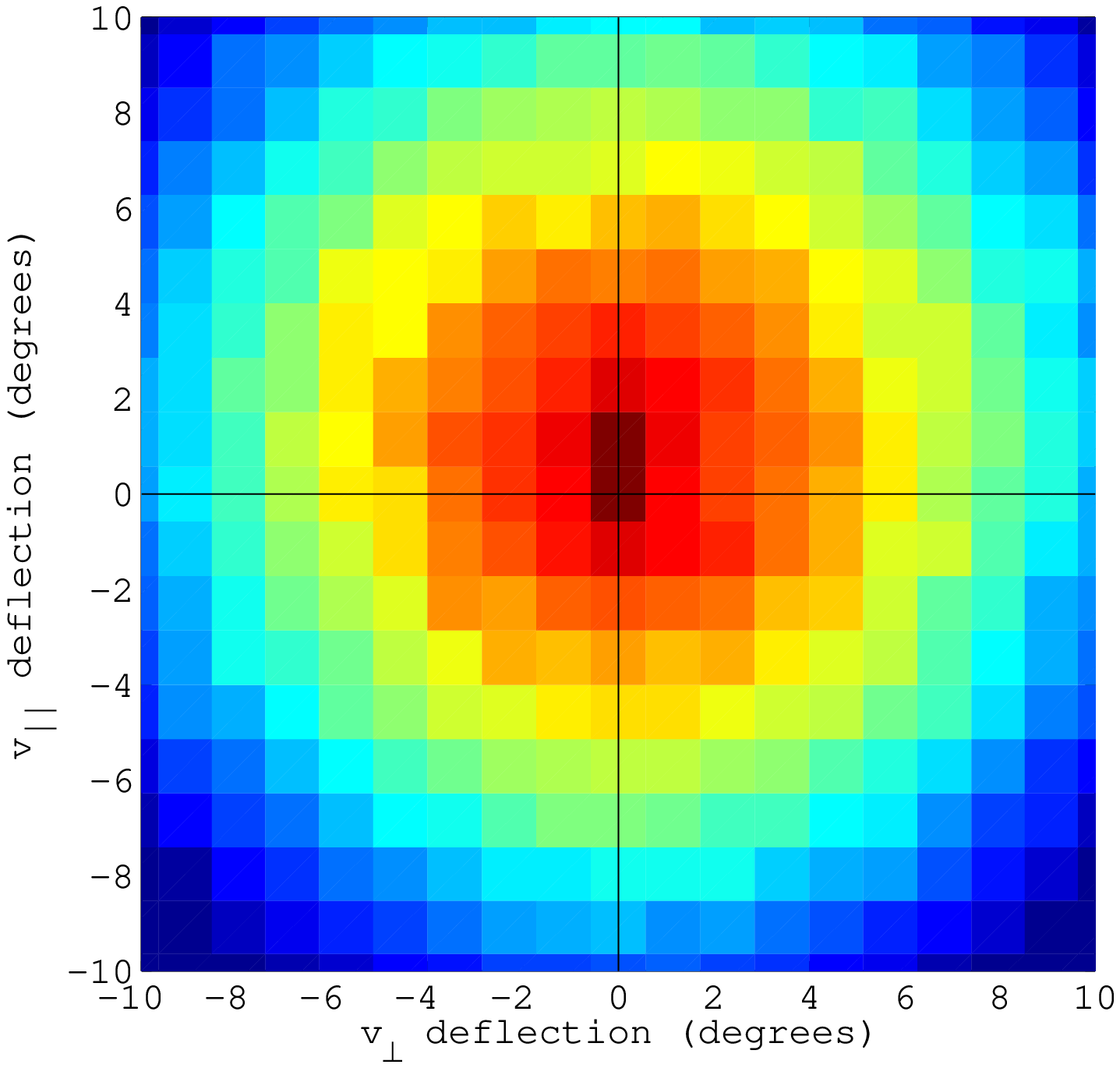}\hspace{0.5mm}
\includegraphics[width=0.25\textwidth]{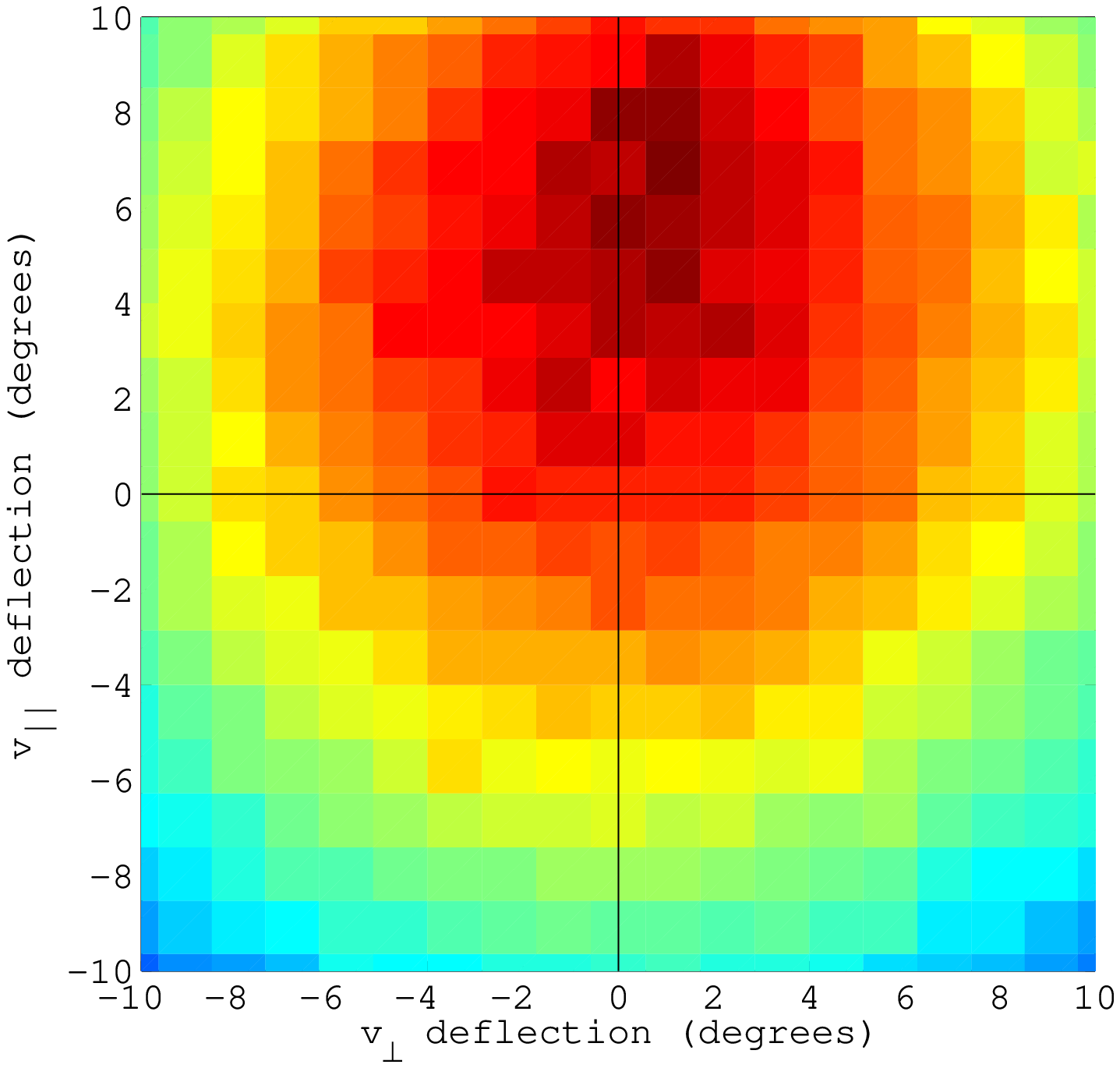}\hspace{0.5mm}
\includegraphics[width=0.25\textwidth]{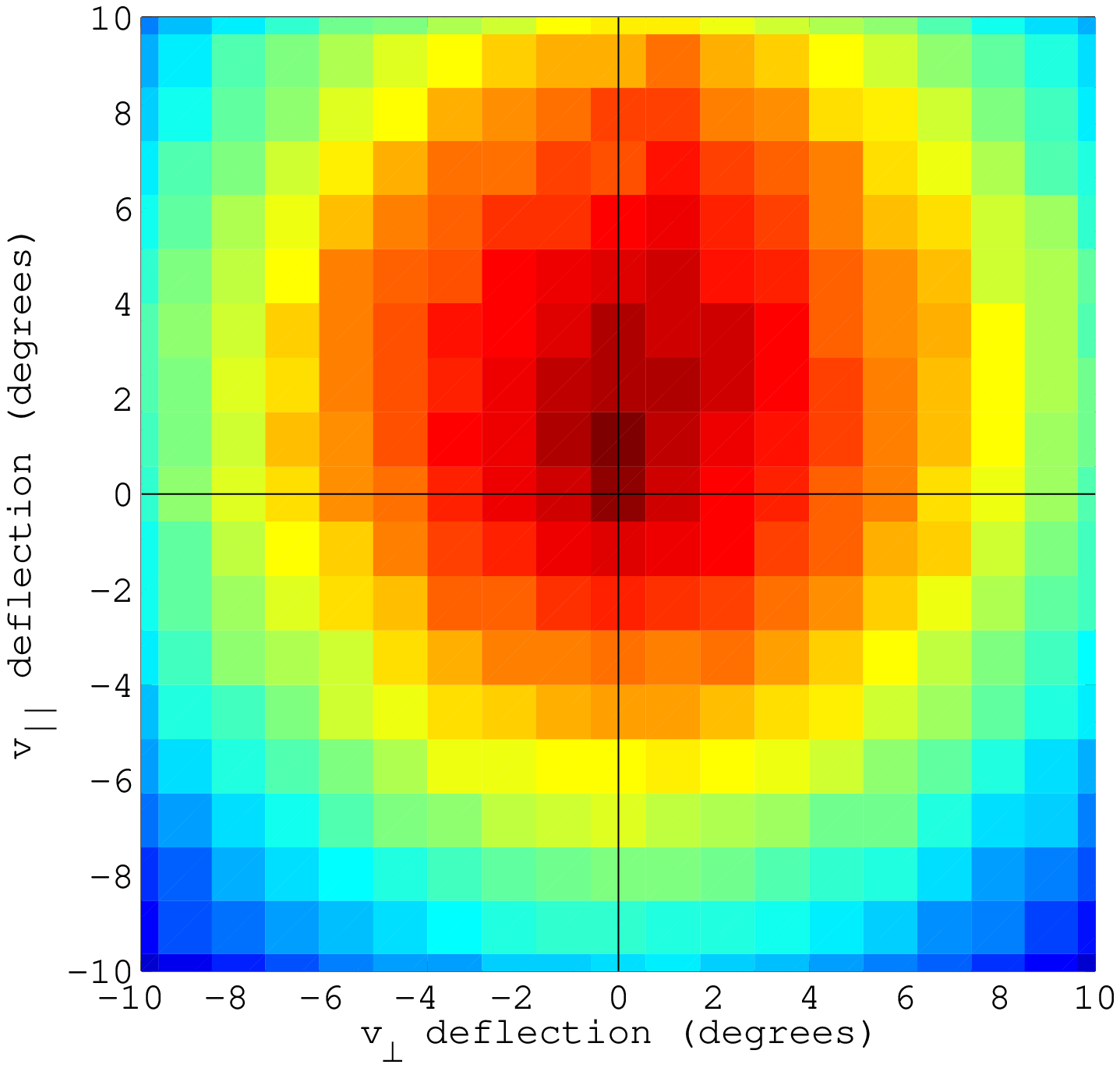}
}
\caption{Two-dimensional distribution of H-atom deflections from $\mathbf{V}_\infty$ in the plane perpendicular to
  the LISM $B$-$V$ plane (the interstellar perspective).
Left are primary interstellar H-atoms, middle are secondary
  (i.~e. last charge-exchange occurred in the outer heliosheath), while
  on the right is the combined distribution. The normal is determined by the vector product
  $\mathbf{r}_\mathrm{H}\times\mathbf{r}_\mathrm{He}$. The darkest red color corresponds to the particle count
  twice larger than that of the darkest blue color.
}
  \label{Jacob}
\end{figure*}

One can try to increase the ISMF intensity to overcome the symmetrizing effect of charge exchange. However, there will be an internal
inconsistency in such an attempt. The main reason for that is that an ISMF-induced asymmetry of the TS and the HP is accompanied by
a more asymmetric distribution of H$^+$ in the outer heliosheath, which in turn results in larger deflection of neutral H from its original
orientation in the LISM (see Pogorelov et al.~2007).
To quantify this effect, we run our kinetic neutral-atom code and collect statistics on the
H-atom velocity distribution in the SW. We record the
deflection from $\mathbf{V}_\infty$ of all H-atoms within a 45-degree cone about $\mathbf{V}_\infty$ out to 80 AU, both
in the $B$-$V$ plane and perpendicular to it, thus creating a two dimensional distribution of deflections.
In Fig.~\ref{Jacob} we show these for
primary (population 0) LISM H-atoms (left), secondary (population 1) H-atoms (middle), and the total (weighted)
distribution (right) in the plane perpendicular to $\mathbf{V}_\infty$.


Although the primary LISM distribution starts
out as Maxwellian, its interaction with the heliosphere results in a
distribution of deflections that is obviously not isotropic. This is
because charge-exchange losses may preferentially
cull a particular part of the distribution, due to asymmetric
plasma flow and the dependence of the charge-exchange rate on the
relative plasma flow speed. Secondary H-atoms (and the combined
distribution, by extension) are clearly not isotropic, and the mean of
the distribution does not coincide with its center, making it more
difficult to quantify the overall deflection.
We find that the average deflection of primary neutrals is about $1.8^\circ$
in the $B$-$V$ plane and $-0.18^\circ$ perpendicular to this plane. For
secondary neutrals, the corresponding values are 4.7$^\circ$ and 0.15$^\circ$,
while for the combined population these are 3.8$^\circ$ and 0.05$^\circ$.  We point
out, however, that the peaks of the distributions are not at these
locations. Instead, the primary population shows a peak close to zero
deflection and the most common deflection of the secondary neutrals is around 7$^\circ$
in the $B$-$V$ plane and 1$^\circ$ out of it.
It appears in this particular example, that the average deflection takes place almost entirely
in the $B$-$V$ plane. Thus, the actual angle between the $B$-$V$ plane and the HDP is determined by the accuracy
of measuring the H-flow direction in the \textit{SOHO} SWAN experiment.
Although we assume that the $B$-$V$ plane is parallel to the average HDP, an additional deflection of the order of
$\pm 1^\circ$ perpendicular to the average HDP cannot be excluded. This gives us an estimate for the angle between
the HDP and $B$-$V$ plane $\arctan{(\tan{1^\circ}/\tan{4^\circ})}\approx 15^\circ$.

\section{Sources of the 2--3 kHz radio emission}
\begin{figure}
\epsscale{0.9}
\centerline{
\plotone{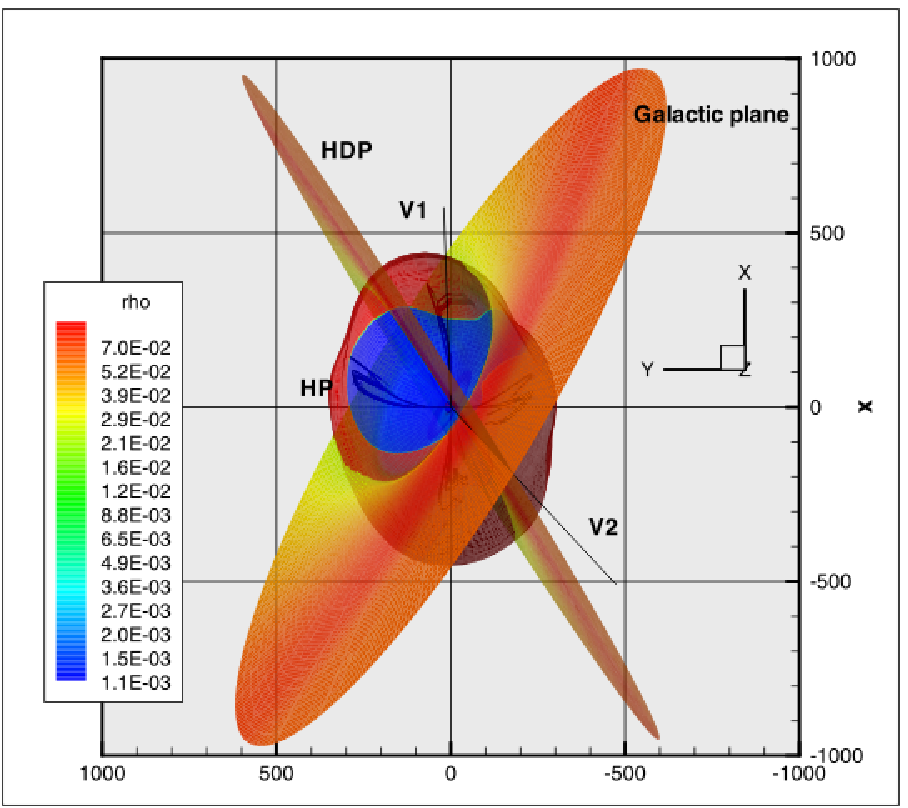}
}
\caption{Frontal view of the HP, hydrogen deflection plane, Galactic plane, \textit{Voyager}~1 and~2 trajectories. Blue and red colors
on the HP surface correspond to $B_R<0$ and $B_R>0$, respectively. The HP is clearly asymmetric with respect to the $B$-$V$ plane.}
\label{front_view}
\end{figure}

Radio emission in the 2--3 kHz range is thought to be generated when a global merged interaction region (GMIR) enters
the outer heliosheath, where plasma is primed with an enhanced level of superthermal electrons \citep{Cairns}.
The origin of the hot electrons is due to their energization by lower-hybrid waves generated by pick-up ions,
created from hot secondary neutrals born in the inner heliosheath. These pick-up ions have a ring-beam distribution.
Since ring-beam driven lower hybrid waves propagate almost perpendicularly to the magnetic field vector $\mathbf{B}$ in the outer heliosheath,
\citet{Gurnett06} suggested that regions which satisfy this property ahead of GMIRs might preferentially radiate in the
2--3 kHz range. If $\mathbf{B}_\infty$ belongs to the observed
HDP, it is then likely that radio emission sources will be distributed approximately in the direction perpendicular to it \citep{Pogo07}.
If one assumes, as was done by \citet{Opher07}, that a GMIR is spherical initially and preserves
its sphericity after crossing the TS and the HP,
then the regions on the HP and in the outer heliosheath where the radial component $B_R$ of the magnetic field is
equal to zero might be candidate regions for radio emission.
The HP shown in Fig.~\ref{front_view} is painted with two colors, red and blue, which correspond to $B_R>0$ and
$B_R<0$, respectively. The boundary between them is where $B_R=0$. Here, like \citet{Pogo07},
we see that there is a
possible-radio-emission band on the HP surface which is nearly perpendicular to the $B$-$V$ plane.
Note that V1 detected radio emission sources
distributed along the Galactic plane, with additional sources not aligned with the Galactic plane and
called ``ambiguous'' by \citet{Kurth03}, who stated that the presence of two different locations at a time is unlikely.
One can see, however, that the approach based on determining regions where $B_R=0$ can exhibit additional bands
with different orientations. It is interesting that if we choose a $B$-$V$ plane parallel to the Galactic plane,
the distribution of magnetic field on the
HP surface will be close to the mirror reflection of that shown in Fig.~\ref{front_view}. Then, we will still have a band tilted at a small angle
to the Galactic plane. That is, the distribution of $B_R=0$ on the HP cannot by itself determine a plane containing $\mathbf{B}_\infty$.
This can only be done by directly correlating the choice of the $B$-$V$ plane with the observed deflection of H-atom from its original direction in
the LISM, as was done in the previous section. To use radio emission data as an additional constraint on the orientation of $\mathbf{B}_\infty$,
one would need to use MHD-kinetic simulations to combine the condition $B_n=0$, where $B_n$ is the magnetic field component
normal to a (nonspherical) GMIR shock surface, with the physical conditions for generation of radio waves.
As the angle between the HDP and the $B$-$V$ plane can be as large as $15^\circ$, the ISMF orientation can be adjusted
to meet radio emission observations.

\section{Conclusions}
New \textit{in situ} and remote observations give us a unique opportunity to constrain previously undetermined properties of the LISM and
of the ISMF threading it. We applied our MHD-kinetic model to determine the deflection of H-atoms from their original direction
in the unperturbed LISM. We showed that the absence of a heliospheric symmetry plane, which is due to the presence of IMF and,
very likely, the regions of slow and fast SW, can introduce an angle between the $B$-$V$ plane and the resulting HDP.
However, this angle is smaller than a possible $15^\circ$-angle between these planes determined by the accuracy
of measuring the H-atom flow direction. This factor, as well as physical conditions for 2--3 kHz radio emission,
can be used to adjust the orientation of the ISMF in such a way that the distribution of radio emission sources observed
by the \textit{Voyager} spacecraft will be approximately aligned with the Galactic plane. The angle of about $30^\circ$ between
$\mathbf{B}_\infty$ and $\mathbf{V}_\infty$ is sufficient to introduce a $4^\circ$-degree average deflection of the H-atom flow from
its original direction in the unperturbed LISM.

\acknowledgements
This work was supported by NASA grants NNG05GD45G, NNG06GD48G, and NNX07AH18G and NSF award ATM-0296114.
Calculations were performed on supercomputers Fujitsu Primepower HPC2500, by
agreement with the Solar-Terrestrial Environment Laboratory of Nagoya University, Columbia at NASA Ames Research Center
(award SMD-06-0167), and IBM Data Star (awards ATM-070011 and MCA07S033) in the San Diego Supercomputer Center.

\clearpage

\clearpage

\clearpage

\clearpage
\end{document}